\documentclass{emulateapj}

\begin{document} 
 
\shorttitle{GLARE Line Emitters at $z\approx6$} 
\shortauthors{E.~R.~Stanway et al.} 
 
\title{Three Lyman-$\alpha$ Emitters at $z\approx6$: Early GMOS/Gemini Data from the GLARE Project} 
 
\author{Elizabeth R.~Stanway\altaffilmark{1}, Karl Glazebrook\altaffilmark{2}, Andrew~J.~Bunker\altaffilmark{1}, Roberto G. Abraham\altaffilmark{3}, Isobel Hook\altaffilmark{4}, 
James Rhoads\altaffilmark{5},  
Patrick J.~McCarthy\altaffilmark{6}, 
Brian Boyle\altaffilmark{7}, Matthew Colless\altaffilmark{8},  
David Crampton\altaffilmark{9},  
Warrick Couch\altaffilmark{10},  
Inger J{\o}rgensen\altaffilmark{11},  
Sangeeta Malhotra\altaffilmark{5},  
Rick Murowinski\altaffilmark{9},  
Kathy Roth\altaffilmark{11},  
Sandra Savaglio\altaffilmark{2},  
Zlatan Tsvetanov\altaffilmark{2} 
}

\altaffiltext{1}{Institute of Astronomy, University of Cambridge, Madingley Road, Cambridge,  CB3\,0HA, U.K., email: ers@ast.cam.ac.uk,bunker@ast.cam.ac.uk} 
\altaffiltext{2}{Department of Physics \& Astronomy,  Johns Hopkins University, 
  3400 North Charles Street, Baltimore, MD 21218-2686, email: kgb@pha.jhu.edu, savaglio@tarkus.pha.jhu.edu, zlatan@pha.jhu.edu} 
\altaffiltext{3}{Department of Astronomy \& Astrophysics,  University of Toronto, 60 St. George Street, Toronto, ON, M5S~3H8, Canada, email: abraham@astro.utoronto.ca} 
\altaffiltext{4}{Department of Astrophysics, Nuclear \& Astrophysics Laboratory, 
 Oxford University, Keble Road, Oxford OX1 3RH, U.K., email: imh@astro.ox.ac.uk} 
\altaffiltext{5}{Space Telescope Science Institute, 3700 San Martin Drive, Baltimore, MD 21218, email: rhoads@stsci.edu, san@stsci.edu} 
\altaffiltext{6}{Observatories of the Carnegie Institute of Washington, Santa Barbara Street, Pasadena, CA~91101, email: pmc2@ociw.edu} 
\altaffiltext{7}{Australia Telescope National Facility, PO Box 76, Epping, NSW 1710, Australia, email: Brian.Boyle@csiro.au} 
\altaffiltext{8}{Anglo-Australian Observatory, PO Box 296, Epping, NSW 1710, Australia, email: colless@aao.gov.au} 
\altaffiltext{9}{Herzberg Institute of Astrophysics, National Research Council, 5071 West Saanich Road, Victoria, 
British Columbia, V9E~2E7, Canada, email: David.Crampton@nrc.ca, Richard.Murowinski@hia.nrc.ca}  
\altaffiltext{10}{School of Physics, The University of New South Wales, Sydney 2052, Australia, email: wjc@edwin.phys.unsw.edu.au} 
\altaffiltext{11}{Gemini Observatory, Hilo, HI 96720, email: ijorgensen@gemini.edu, kroth@gemini.edu} 
 
 
\begin{abstract}  
 
We report spectroscopic detection of three $z \sim 6$ Lyman-$\alpha$
emitting galaxies, in the vicinity of the Hubble Ultra Deep Field,
from the early data of the Gemini Lyman-$\alpha$ at Reionisation Era
(GLARE) project.  Two objects, GLARE\#3001 ($z =5.79$) and GLARE\#3011
($z =5.94$), are new detections and are fainter in $z'$ ($z'_{AB} =
26.37$ and $27.15$) than any Lyman break galaxy previously
detected in Lyman-$\alpha$.  A third object, GLARE\#1042 ($z =5.83$)
has previously been detected in line emission from the ground; we
report here a new spectroscopic continuum detection.  Gemini/GMOS-S
spectra of these objects, obtained using nod \& shuffle, are presented
together with a discussion of their photometric properties.  All three
objects were selected for spectroscopy via the $i$-drop Lyman Break
technique, the two new detections from the GOODS v1.0 imaging data.
The red $i'-z'$ colors and high equivalent widths of these objects
suggest a high-confidence $z>5$ Lyman-$\alpha$ identification of the
emission lines.  This brings the total number of known $z>5$ galaxies
within 9\arcmin\ of the Hubble Ultra Deep Field to four, of which
three are at the same redshift ($z=5.8$ within 2000 km s$^{-1}$),
suggesting the existence of a large-scale structure
at this redshift.
\end{abstract} 
 
\keywords{ 
galaxies: evolution -- 
galaxies: formation -- 
galaxies: staburst -- 
galaxies: individual: SBM03\#1, SiD002 -- 
galaxies: high redshift -- 
ultraviolet: galaxies}

\section{Introduction} 
\label{sec:intro} 
 
The detection and analysis of objects at very high redshift is a
challenging but rapidly advancing field.  In recent months increasing
numbers of galaxies have been found at redshifts of $z\approx6$, as a
result both of narrow-band selection (e.g. Rhoads et al.\ 2003,
Taniguchi et al.\ 2003, Hu et al.\ 2003) and the use of Lyman break
methods (e.g. Dickinson et al.\ 2003, Stanway, Bunker \& McMahon 2003
hereafter SBM03). However, until now, spectroscopic confirmation of
very high redshift candidates has only been possible for objects at
the bright end of the galaxy luminosity function.
 
Spectroscopic surveys for fainter objects are essential to 
shape our understanding of the universe at these redshifts and, with 
modern instrumentation and new techniques for faint-object 
spectroscopy, such surveys are now possible.  The Gemini 
Lyman-$\alpha$ at Reionisation Era (GLARE) survey, described in this 
paper, aims to obtain extremely deep spectra of very faint high 
redshift starforming galaxies, selected via $i-z$ color from 
the HST/ACS Ultra Deep Field (UDF, Beckwith et 
al.\ 2003) and  Great Observatories Origins Deep Survey 
(GOODS, Giavalisco et al.\ 2003) data, using the GMOS spectrograph on Gemini 
South. 
 
In this Letter we present spectra and fluxes\footnote{We adopt 
the following cosmology: a flat Universe 
with $\Omega_{\Lambda}=0.7$, $\Omega_{M}=0.3$ and $H_{0}=70 h_{70} 
{\rm km\,s}^{-1}\,{\rm Mpc}^{-1}$. All magnitudes are 
quoted in the AB system (Oke \& Gunn 1983).}
of three Lyman-$\alpha$ emitters, two of which are new, 
identified in the early observation data of the GLARE project.

\section{Gemini GMOS Observations} 
\label{sec:GMOSobs} 
 
\def\th{\hbox{$^{\rm th}$}} The spectra described in this paper were 
obtained as part of the Gemini South program GS--2003B-Q-7 using the 
Gemini Multi-Object Spectrograph (GMOS, Hook et al.\ 2003) targeting 
sources in the region of the Hubble UDF \citep{Newsletter}.
 
The goal of program GS--2003B-Q-7 was to obtain a total of 100 hours
exposure on this UDF mask. This letter concerns the first 7.5 hours of
data.  Observations were made with the R150 grating at 9000\AA\
central wavelength and a custom made 7800\AA\ longpass filter
(`RG780') giving a spectral range from 7800\AA\ up to the CCD cutoff
(about 10,000\AA) and a spectroscopic resolution of 15\AA\ (4 pixels
FWHM at 3.5 \AA/pix) with 0\farcs7 wide slits, a resolving power of
$\lambda\,/\,\Delta\lambda_{\rm FWHM}\approx550$. As the seeing disk
($<0$\farcs5 FWHM) was smaller than the slit width (0\farcs7), the
true resolution is somewhat better for a source which does not fill
the slit.
 
The data was taken using the `nod \& shuffle' (N\&S) observing mode
(Cuillandre et al.\ 1994, Glazebrook \& Bland-Hawthorne 2001, Abraham
et al. 2004 - hereafter GDDS1). Our N\&S setup and observational
scheme follows that in GDDS1 except that the slits were
2\farcs5 long with a 1\farcs5 nod. Data reduction also follows GDDS1
(Appendix B and C).  Fifteen 1800\,s frames were obtained 
in queue observing during November 2003 (19\th, 24\th,
28\th\ and 30\th) during conditions of 0.4--0.5 arcsec seeing and high
transparency.  Relative flux
calibration was performed using observations of a standard star, the
absolute flux calibration was done by normalizing the spectra of the
mask alignment stars to their $z'$-band photometric fluxes. Sky lines were used for wavelength calibration.
 
1D and 2D sky-subtracted spectra were inspected for the presence of 
emission lines. An advantage of the N\&S data reduction technique is that 
lines appear in a positive-negative dipole pattern in the 2D images
making them easy to distinguish from 
residual CCD defects and other non-astrophysical effects.

\section{HST/ACS Photometry and Candidate Selection} 
\label{sec:ACSimages} 

Spectroscopic candidates for this program were selected using the
$i$-drop Lyman break technique (see SBM03 or Dickinson et al.\ 2003).
This photometric selection method has been used by a
number of authors to identify $z>5$ galaxies (e.g. Bremer et al.\
2004, Lehnert \& Bremer 2003 at $z\sim5$, Dickinson et al.\ 2003,
SBM03 at $z\sim6$). In each of these surveys the Lyman break spectral
feature is detected by means of a colour-cut criterion. At $z\sim6$
the lyman break passes through the $z'$ filter and hence a cut
is usually placed in the range $1.3<i'-z'<1.5$ with the exact value
affecting the survey redshift range and low
redshift contamination.

Within the UDF itself, 19 candidates
were drawn from the catalogue of the reddest ($i'-z'$) objects,
pre-released by the UDF team\footnote{http://www.stsci.edu/hst/udf/}
in order to facilitate ground-based spectroscopic follow up and
satisfy $(i'-z')_{\rm AB}>1.5$. The 5.5\arcmin\ GMOS field is bigger
than the UDF, so in the outlying region we selected objects from the
 GOODS-S field.  The GOODS v1.0 data release\footnote{available from
{\tt ftp://archive.stsci.edu/pub/hlsp/goods/v1/}} comprises coadded
imaging from 5 `epochs' of observations, reaching $3\sigma$ magnitude
limits $v_{{\rm lim}}=29.44$, $i'_{{\rm lim}}=28.83$ and $z'_{{\rm
lim}}=28.52$. We selected candidates using $(i'-z')_{\rm AB}>1.3$ and
$z'_{\rm AB}<27.5$\,mag and identified 18 objects in the slitmask area
of which 10 were placed on the mask. The faint end of this selection
reaches the magnitude limit in the $i$-band. The rest of the mask area
was used for a blank sky survey. Of the objects presented in this
paper \#1042 was identified as a candidate in both catalogs, while
\#3001 and \#3011 lie outside the UDF field and were selected from the
GOODS v1.0 data.

\begin{figure} 
\epsscale{0.8} 
\plotone{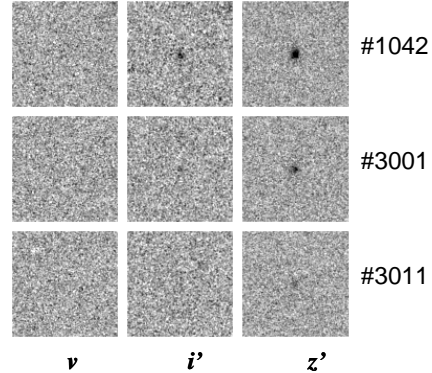} 
\caption{GOODS-S imaging of GLARE\#1042, GLARE\#3001 and GLARE\#3011 in $v$(F606W), $i'$(F775W) and $z'$(F850LP). Boxes are 3\arcsec\ to a side (17 $\,h_{70}^{-1}\,$kpc at $z=5.8$). North is up.} 
\label{fig:images} 
\end{figure}

Table \ref{tab:photspec} presents the broadband photometric properties of 
these objects. Magnitudes are measured in 
a 0\farcs3 diameter aperture and an aperture correction of -0.32mag 
(measured from point sources on the images) is applied to obtain a 
total magnitude for these compact sources. Postage stamp images in 
the $v$, $i'$ and $z'$ bands are shown in figure 
\ref{fig:images}. Since \#1042 lies within the UDF more accurate 
photometry and morphological information should soon be available 
for this object. 
All three objects are spatially resolved with half-light radii in the range 
0\farcs09-0\farcs14 (cf. stellar value of 0\farcs05), well detected in the 
$z'$ band, faintly detected (or in the case of \#3011 
undetected at $3\sigma$) in $i'$ and undetected at 3$\sigma$ in the 
$v$ band. 

 
\section{Results} 
\label{sec:results} 
 
Single isolated emission lines have been identified for each of three 
objects.  GLARE\#1042 (J2000 3$^{\rm h}$32$^{\rm m}$40\fs0 
$-$27\degr48\arcmin15\farcs0), is an $i$-drop initially selected by 
SBM03 and already confirmed as a z=5.83 Lyman-$\alpha$ emitter 
(Stanway et al.\ 2003, Dickinson et al.\ 2003).  GLARE\#3001 (3$^{\rm 
h}$32$^{\rm m}$46\fs0 $-$27\degr49\arcmin29\farcs7) and \#3011 
(3$^{\rm h}$32$^{\rm m}$43\fs2 $-$27\degr45\arcmin17\farcs6) are new.  
Their properties are given in table 
\ref{tab:photspec}. All three lines are almost 
unresolved given our dispersion, although \#1042 is known to be highly 
asymmetric from published spectra (Stanway et al.\ 2003, 
Dickinson et al.\ 2003). 
 
\begin{figure} 
\epsscale{1.} 
\plotone{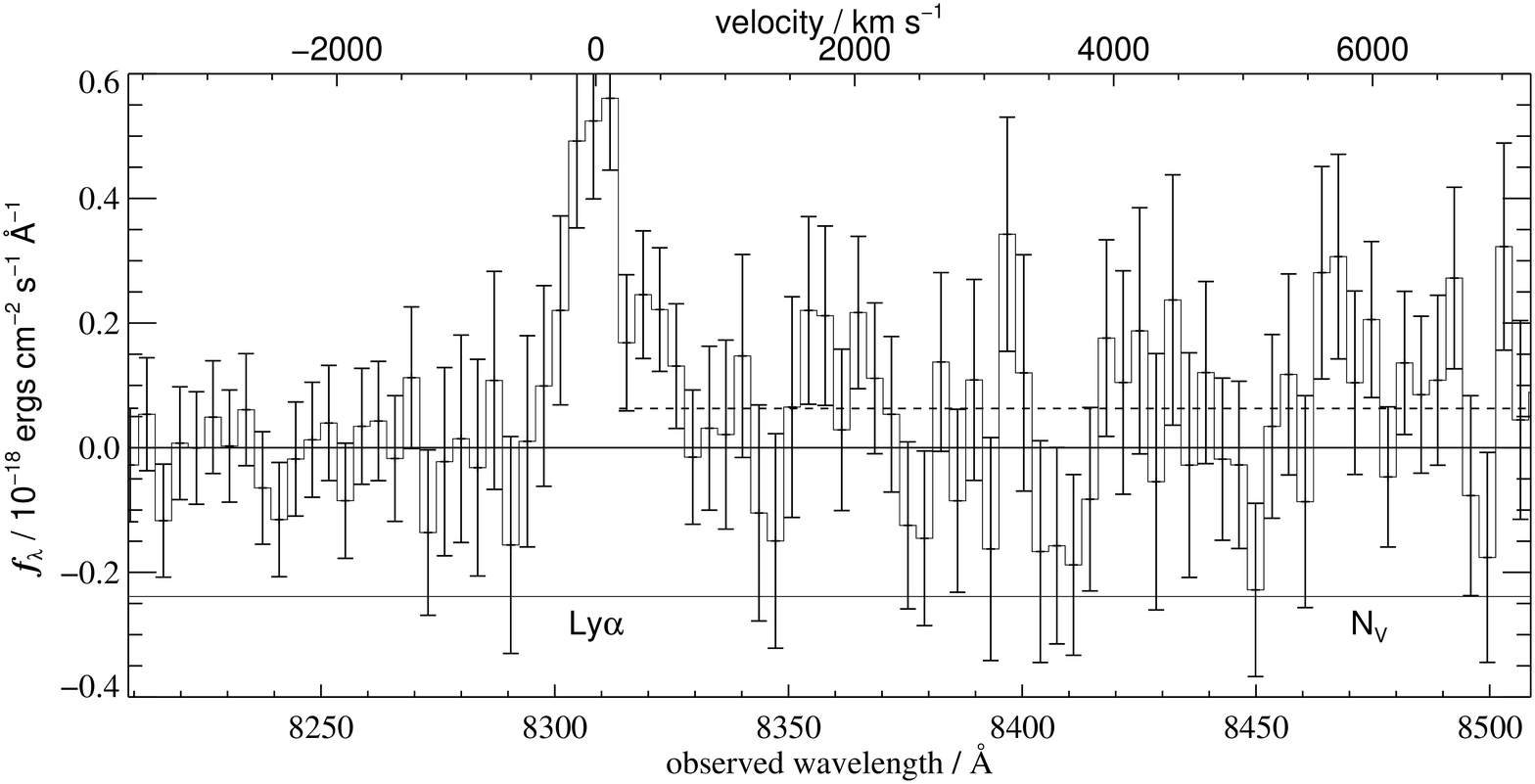}
\epsscale{1.} 
\plotone{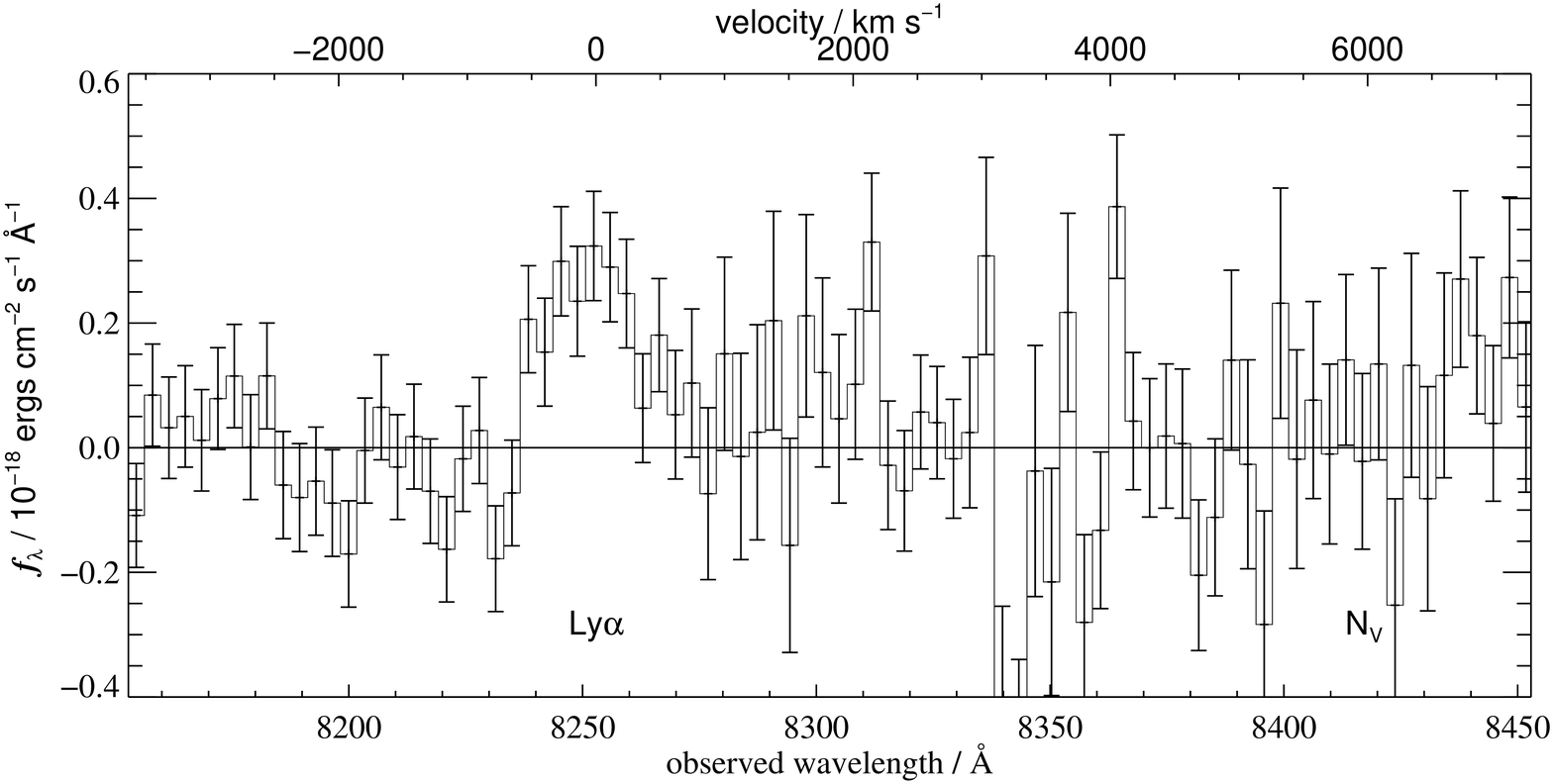} 
\epsscale{1.} 
\plotone{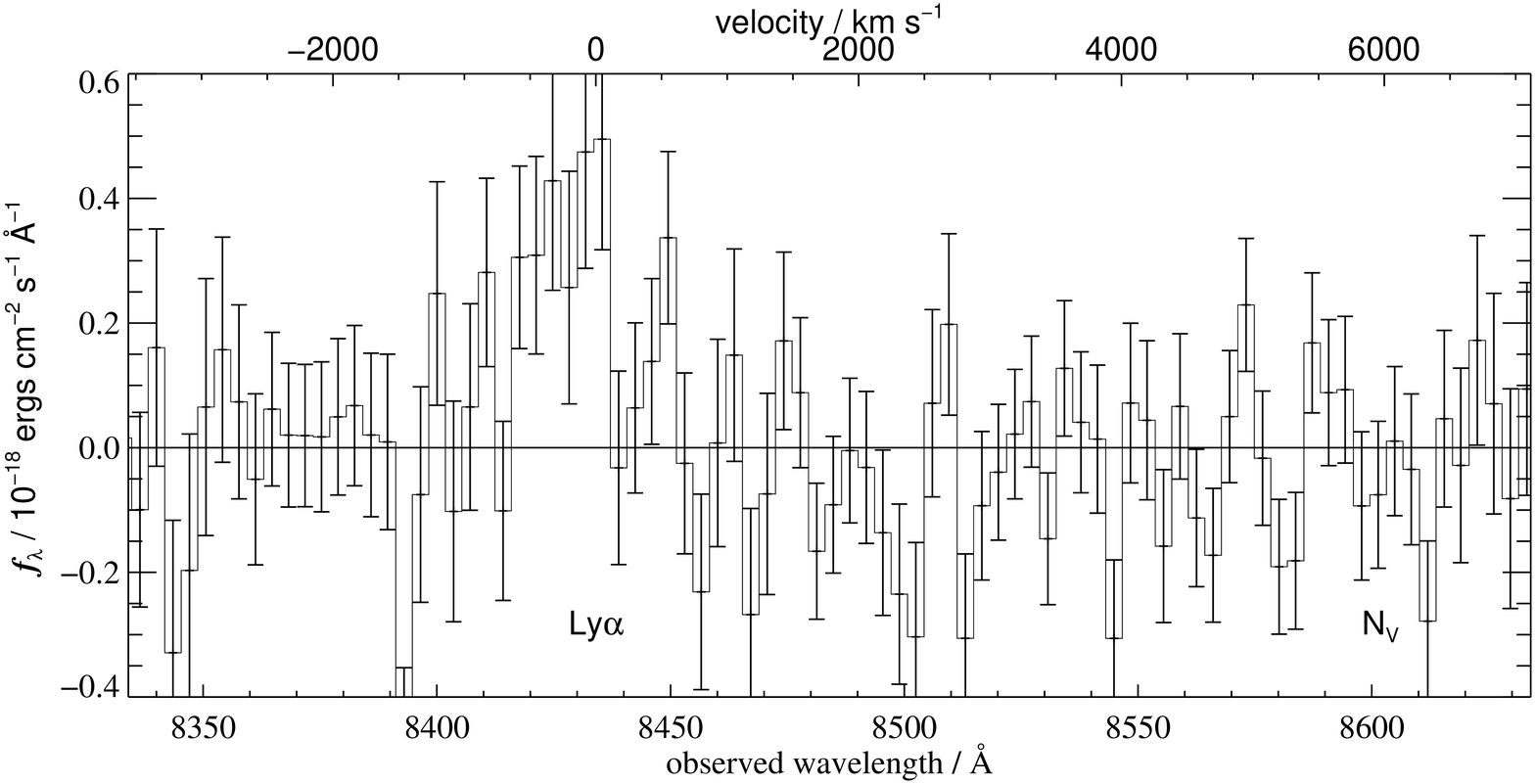} 

\caption{\scriptsize Unbinned GMOS-S spectra of GLARE\#1042,
GLARE\#3001 and GLARE\#3011 around
the Lyman-$\alpha$ candidate line emission. 
The continuum level is indicated on \#1042}
\label{fig:spectra} 
\end{figure}

We also detect the continuum break in the spectrum of GLARE\#1042
although at low signal-to-noise (Flux density $\approx 0.6 \times
10^{-19}$ ergs cm$^{-2}$ s$^{-1}$ \AA$^{-1}$). This is consistent with
the level of continuum reported in the low-resolution ACS slitless
spectrum of Dickinson et al.\ and also with the broadband colours of
this object. A good continuum spectrum of this object should be
obtained if GLARE does achieve a 100 hour total exposure. The spectra
of the candidates are shown in figure \ref{fig:spectra}.


\thispagestyle{empty}

\begin{deluxetable*}{lcccccccccc} 
\tablecaption{Summary of Spectroscopic and Photometric Properties \label{tab:photspec}} 
\tablewidth{0pt} 
%

\tablehead{ 
ID \# & $z$ &Peak & Flux(1) & Flux(2) & FWHM  & $z'$ & $i'-z'$ & R$_{h}$ & SFR$_{UV}$  \\ &  & \AA\ & ergs cm$^{-2}$ s$^{-1}$ & ergs cm$^{-2}$ s$^{-1}$ & \AA\ & & & ($z'$) & $h^{-2}_{70}\,M_{\odot}\,{\rm 
yr}^{-1}$ } 
\startdata 
 
1042 & 5.83 & 8309.1  & $0.97\times10^{-17}$  & $0.91\times10^{-17}\ (12\sigma)$ & 16.3 &   25.48 $\pm$ 0.03  & 1.48 $\pm$ 0.09   & 0\farcs09 & 15  \\ 
3001 & 5.79 & 8252.7  & $0.79\times10^{-17}$  & $0.76\times10^{-17}\ (12\sigma)$ & 20.5 &   26.37 $\pm$ 0.06  & 1.66 $\pm$ 0.20   & 0\farcs14 & 6.0 \\ 
3011 & 5.94 & 8434.0  & $1.0\times10^{-17}$   & $ 1.0\times10^{-17}\ (25\sigma)$ & 24.5 &   27.15 $\pm$ 0.12  & $>1.68\ (3\sigma)$& 0\farcs13 & 1.7 \\

\enddata 
\tablecomments{The spectroscopic and photometric properties of the three emission line candidates. Wavelengths are measured in air. Error on the fluxes due to slit losses and photon noise is $\approx20\%$. Flux(1) is taken between zero power points, Flux(2) by fitting a gaussian line profile to the data.  The FWHM of this fit is shown in the sixth column (spectral resolution $\approx16.5$\AA). 
All objects are undetected at 3$\sigma$ in $v$ ($v > 29.4$) and also resolved in the $z'$ band (stellar half-light radius R$_{h}$=0\farcs05). Star Formation Rates are calculated using the UV Flux-SFR relation of Madau, Pozzetti \& Dickinson (1998) having removed line contamination (and assuming $\beta=-1.1$).} 
\end{deluxetable*} 
 
 
\section{Interpretation} 
\label{sec:interp} 
 
\subsection{The Redshifts of GLARE\#1042, \#3001 and \#3011} 
\label{subsec:redshifts}
The most plausible line identifications given the red $i-z$ selection
are Lyman-$\alpha$ at 
$z=5.83$, $z=5.79$ and $z=5.94$ for \#1042, \#3001 and \#3011 
respectively. H$\beta$\,$\lambda$\,4861.3\,\AA\ 
or [O{\scriptsize~III}]\,$\lambda\lambda$\,5006.8,4958.9\,\AA\ at 
$z\approx 0.7$ or H$\alpha$\,$\lambda$\,6562.8\,\AA\ 
 at $z\approx0.25$  are ruled out be the absence of nearby lines and
because galaxies at these redshifts do not have strong $i-z$
continuum breaks.
 
An unresolved
[O{\scriptsize~II}]\,$\lambda\lambda$\,3726.1,3728.9\,\AA\ doublet at
redshifts $z\approx1.2$ is a possibility for \#3001 and \#3011 (\#1042
is already known to be strongly asymmetric ruling this out, Stanway et
al.\ 2003).  The best evidence against this is the red $i-z$
color. The reddest possible color from an early-type SED at $z=1.2$ is
$i'-z'=1.2$ (SBM03) and our two new objects are redder than this at
95\% confidence. An [O{\scriptsize~II}] identification would also
imply rest-frame equivalent widths of 50--250\AA\ (see
Section~\ref{subsec:EW}) higher than is normally seen in $z\sim 1$
[O{\scriptsize~II}] emitters (Hammer et al. 1997).


\subsection{Equivalent Widths} 
\label{subsec:EW}

  The emission line equivalent widths can  be calculated from their
$z'$ band photometry and line fluxes.  However, a subtlety
arises as the  emission lines for these objects contribute
significantly to their  continuum flux in both the $z'$ and $i'$
filters; the calculated  equivalent widths are sensitive to the
detailed shape of the filter  transmission curve.  Accounting for
these effects, we calculate  equivalent widths $W_{\rm rest}^{\rm
Ly\alpha}=[20, 30, 100]\pm10$\AA\  for [\#1042, \#3001, \#3011]
respectively, assuming that there is negligible flux
short-ward of the Lyman-$\alpha$ line due  to absorption by the
Lyman-$\alpha$ forest.  From stellar synthesis  models of
star-forming regions (e.g., Charlot \& Fall 1993), the  theoretical
Lyman-$\alpha$ equivalent width for a young region of  active star
formation is $W_{\rm rest}^{\rm Ly\alpha}\approx  100-200$\,\AA.
However, observed Lyman-$\alpha$ emission from star-forming galaxies
is generally weaker in continuum-selected samples at lower
redshifts, which typically have $W_{\rm rest}=5-30$\,\AA\ (e.g.,
Steidel et al.\ 1996), or even in
absorption. Our $z \simeq 6$ sample has a comparable selection but the
equivalent widths of \#3011 is notably larger. We note that $z \simeq 6$
narrow-band selected samples also produce higher equivalent widths with a median of 200\AA\ (Malhotra \& Rhoads 2002, Hu et al.\ 2003).
 
\subsection{AGN or Starburst?} 
\label{subsec:AGNorStarburst} 
 
Careful inspection of the spectra reveal no other lines.
The only other significant line in our range 
is the high-ionization rest-UV  
N{\scriptsize~V}\,$\lambda\,1240$\,\AA, which is 
usually prominent in active galactic nuclei (AGN).  
Our flux limits at $\lambda_{\rm rest}1240$\,\AA\ are $f < [2.3, 2.3,
 1.8] \times 10^{-18}\,{\rm ergs\,cm}^{-2}\,{\rm s}^{-1}$
 ($3\,\sigma$), for a 15\AA\ FWHM (unresolved) line. This gives lower
 limits of $f({\rm Ly\alpha})/f({\rm N{\scriptsize~V}})> [4.3,3.5,5.6]
 $ ($3\,\sigma$).  Typical line ratios from composite QSO spectra are
 $f({\rm Ly\alpha})/f({\rm N{\scriptsize~V}})=4.0$ (Osterbrock
 1989). Although our constraints are admittedly weak, the
 non-detection of N{\scriptsize~V}\,1240\,\AA\ favours the
 interpretation that the Lyman-$\alpha$ arises from the Lyman
 continuum flux produced by OB stars, rather than the harder UV
 spectrum of a QSO.
 
 
None of the objects in this paper correspond to X-ray sources in the 
Chandra Deep Field South 1Ms catalog which covers this region 
(Giacconi et al.\ 2003). This allows us to place a limit on their X-ray 
fluxes in the $0.5-2$\,keV (soft) and $2-8$\,keV (hard) bands at 
$\sim5.5\times10^{-17}$ and $\sim4.5\times10^{-16} h^{-2}_{70}\,$ ergs 
cm$^{-2}$ s$^{-1}$ respectively. The hard X-ray flux limits rule out
a type II QSO like  CDF-S 202 (Norman et al 2002) but not
CXO-52 (Stern et al.\ 2002). The soft X-ray limits would permit low-luminosity
Seyferts and QSOs (Malhotra et al. 2003),  but the line emission 
we see is not broad.
 
 
Given that an AGN interpretation for the observed emission lines is 
unlikely, the rest frame-UV spectra of these objects is likely to be 
dominated by star formation. The relation between the flux density in 
the rest-UV around 1500\,\AA\ and the star formation rate 
(${\rm SFR}$ in $M_{\odot}\,{\rm yr}^{-1}$) is 
given by $L_{\rm UV}=8\times 10^{27} {\rm SFR}\,{\rm 
ergs\,s^{-1}\,Hz^{-1}}$ (Madau, Pozzetti \& Dickinson 1998) for a 
Salpeter (1955) stellar initial mass function (IMF) with 
$0.1\,M_{\odot}<M^{*}<125\,M_{\odot}$. The rest frame-UV 
at $z\approx6$ is redshifted into the $z'$ photometric band. 
The inferred SFR for the objects in this paper, given their $z'$ 
magnitudes, line contamination and identified redshifts, are shown in 
table \ref{tab:photspec}.

\subsection{Overdensity at z=5.8?} 
\label{subsec:overdensity} 
 
The redshift of new Lyman-alpha emitter GLARE\#3001 is very similar to 
that of Lyman-alpha emitter SBM03\#3 reported by Bunker et al.\ (2003), 
and is also close to that of the re-confirmed $z=5.83$ object 
(variously GLARE\#1042, SBM03\#1, SiD002, reported in this paper, 
Stanway et al.\ 2003, Dickinson et al.\ 2003).
The $i$-drop selection procedure used to identify candidate objects
for spectroscopy is in principle sensitive to objects in the redshift
range $5.6<z<7.0$ although the sharp fall off in transmission of the
$z'$ filter effectively restricts detection of objects to a smaller
redshift range which varies with absolute magnitude ($5.8<z<6.5$ for
an $L^*_{z=3}$ Lyman Break galaxy, see SBM03). The line centers of
three of the four Lyman-$\alpha$ emitters detected in the GOODS-S
field are separated by 2000\,km\,s$^{-1}$ (only $\approx5$\% of the
survey redshift range).  The maximum angular separation of the the
three objects (GLARE\#1042, \#3001, SBM03\#3) is less than
11.5\,arcmin, corresponding to a projected separation of only
$4.0\,h_{70}^{-1}$\,Mpc. This suggests the interesting possibility of
an overdensity of Lyman-$\alpha$ emitters at $z=5.8$ in this field,
perhaps similar to that observed at $z=3.1$ in SSA22 by Steidel et
al.\ (2001).

To quantify the statistical likelihood of this happening by chance we
undertook a Monte-Carlo simulation. A population uniformly distributed
in redshift between $5.6<z<7.0$ was considered and the effects of GMOS
throughput (including the RG780 filter) and increased noise due to
skylines on spectroscopic detection, and IGM absorption leading to
incomplete $z'$ filter coverage on Lyman break photometric selection,
were modelled.  The effects of an apparent magnitude limit on the
photometric selection were also calculated, using the measured
luminosity function for Lyman break galaxies at $z=3$ (Steidel et al.\
1999) as an approximation for the currently unknown $z=6$ luminosity
function. Although each of these effects bias detection probability
towards the lower end of the accessible redshift range, the
probability of selecting 3 or more out of 4 galaxies within $\pm$2000
km s$^{-1}$ of $z=5.80$ was still small ($2.1\%$).  This suggests that
we have statistical evidence for a redshift spike at $z=5.8$ in this
field, which augurs well for the future UDF observations.
 
\subsection{Comparison with Previous Work} 
\label{subsec:comp}

The spectroscopic properties of the three galaxies presented in this
work are similar to those of other confirmed high redshift Lyman break
galaxies.  Lehnert \& Bremer (2003) report Ly-$\alpha$ luminosities of
$10^{42}-10^{43}$ ergs s$^{-1}$ for 6 $z=5$ line emitters in a sample
of 12 Lyman break candidates.  The fluxes for our $z=6$ Ly-$\alpha$ emitters
fall comfortably into this range of luminosities and probe
to a comparable depth.  The emission lines equivalent widths of the
GLARE candidates are also comparable to those of other
spectroscopically confirmed Lyman break galaxies,
e.g. at $z=5.8$ $W_{\rm rest}^{\rm Ly\alpha}=20$\AA\ (Bunker et al.\
2003) and $z=6.17$ $W_{\rm rest}^{\rm Ly\alpha}=50$\AA\ (Cuby et
al.\ 2003).

\section{Conclusions} 
\label{sec:conclusions} 
 
In this paper we have presented photometry and spectra for three objects 
with extreme $i'-z'$ colors and line emission which may be Lyman-$\alpha$ at $z\approx 5.8$, observed as part of the GLARE project. 
Our main conclusions can be summarized as follows: 

1) The GLARE project has detected three very high redshift objects 
($z=5.83, 5.79, 5.94$) in the first 7.5 hours of integration time on 
Gemini/GMOS-S. To the best of our knowledge the two new Lyman-$\alpha$
emitters are fainter in $z'$ than any previous 
Lyman break selected objects with a spectroscopic redshift.

2) $i'-z'$ color selection can successfully identify objects lying at these redshifts and magnitudes and the faint end of the galaxy luminosity function at $z=6$ is now
within the reach of 8m telescopes. 

3) We have evidence for an overdensity
of $z=5.8$ objects in a narrow redshift spike in the GOODS-S field;
further observations, for example narrow-band imaging, would be
invaluable in confirming the existence of such a structure.   

\acknowledgements
Based on observations obtained at the Gemini Observatory, which is operated 
by AURA under a cooperative agreement
with the NSF on behalf of the Gemini partnership: NSF (U.S.), PPARC  (U.K.),
NRC (Canada), CONICYT (Chile), ARC
(Australia), CNPq (Brazil) and CONICET (Argentina). KG \& SS
acknowledge generous funding from the David and
Lucille Packard Foundation. 
Also based on observations made with the NASA/ESA Hubble Space Telescope, obtained at the STScI, which is operated by AURA, under NASA contract NAS 5-26555.

\end{document}